\documentclass[11pt]{article}
\usepackage{amsmath,cite}

\newcommand{\ben}{\begin{eqnarray}\displaystyle}
\newcommand{\een}{\end{eqnarray}}
\newcommand{\refb}[1]{(\ref{#1})}

\begin{document}

\thispagestyle{empty}

\begin{flushright}
hep-th/0608132\\
HRI-P-0608003\\
TIFR/TH/06-22
\end{flushright}

\vskip 1.5cm

\begin{center}
{\Large \bf Generalised HyperK\"ahler Manifolds}\\
{\Large\bf in String Theory}

\vspace*{8.0ex}

{\large\rm Bobby Ezhuthachan\footnote{\tt bobby@theory.tifr.res.in}}

\vspace*{1.5ex}

{\it Department of Theoretical Physics}\\
{\it Tata Institute of Fundamental Research}\\ 
{\it Homi Bhabha Road, Mumbai 400005, India}
 
\vspace*{1.5ex}

and\\

\vspace*{1.5ex}

{\large\rm Debashis Ghoshal\footnote{\tt ghoshal@mri.ernet.in}}

\vspace*{1.5ex}

{\it Harish-Chandra Research Institute}\\ 
{\it  Chhatnag Road, Jhusi, Allahabad 211019, India}

\vspace*{12ex}

{\bf Abstract}

\begin{quote}
We discuss the notion of generalised hyperK\"ahler structure
in the context of string theory and discuss examples of this
geometry. 
\end{quote}
\end{center}

\newpage

\setcounter{page}{1}



\section{Introduction}
Strings, being extended, require a generalisation of the geometry of
spacetime that goes beyond the conventional notion. This is at the
heart of novel and beautiful ideas such as mirror symmetry. More
recently there has been a lot of advances in understanding the motion
of a string in the `compact' internal space with various fluxes that
are possible in string theory. This has prompted the investigation of
an appropriate geometrical framework. One such idea is that of the
generalised complex geometry introduced by
Hitchin\cite{HitchinGCY,HitchinGK} and developed further by
Gualtieri\cite{Gualtieri}, in which both the tangent as well as the
cotangent spaces of the manifold are considered together. As a matter
of fact, this occurs naturally in string theory when there is a
magnetic flux of the Neveu-Schwarz (NS) $B$-field.  Demanding
supersymmetry under this condition requries that the left- and
right-moving modes of the string perceive different complex (or
K\"ahler or hyperK\"ahler) structures\cite{GaHuRo}--\cite{HoPa}. In
purely geometrical terms, this turns out to be equivalent to a
generalised complex (or K\"ahler or hyperK\"ahler)
structure\cite{Gualtieri} (see also \cite{KaYi}).  Essentially these
two descriptions are the lagrangian and hamiltonian approaches to the
string sigma model. A number of papers\cite{recone}--\cite{recftn}
have analysed various aspects of generalised geometry in the context
of string theory in recent times (for a recent review, see
\cite{genCrev}).

In this short note, we will elaborate on the generalised hyperK\"ahler
structure alluded to by Hitchin\cite{HitchinGCY,HitchinGK} and
discussed by others\cite{Gualtieri,mathhK} and mention a few examples
of generalised hyperK\"ahler manifolds that are found in string
theory. In view of the comments above, it is clear that these ought to
be alternative descriptions of string backgrounds that can be
equivalently described in terms of left- and right-movers. Our first
example is indeed a reformulation of the familiar Neveu-Schwarz
fivebrane\cite{CaHaSt} in terms of generalised geometry.

It should also be mentioned that string theory allows for fluxes of
various other fields. In particular, there are generalised gauge form
fields of all degrees from Ramond sector. Although some work has been
done, finding a geometrical framework for diverse flux vacua remains a
challenging open problem.

In the following, we briefly recapitulate some essential aspects of
generalised complex and K\"ahler structures. This is then extended to
generalised hyperK\"ahler structure. We provide an example of
generalised hyperK\"ahler geometry from string theory. In the end, we
propose construction of other possible examples.

\bigskip

\noindent{\bf Note Added:} The paper \cite{BredhK}, which appeared
while we have been studying this problem, has some overlap with our
work. However, the focus of \cite{BredhK} is on the analysis of the
worldsheet sigma model and supersymmetry, while we have provided
explicit examples which satisfy these conditions.


\section{Generalised complex and K\"ahler geometry}

A generalised complex structure extends the notion of the usual
complex structure to the sum of the tangent and cotangent bundles
$T{\cal M}\oplus T^*{\cal M}$ of a manifold ${\cal M}$.  

An almost generalised complex structure in an open neighbourhood of
$p\in{\cal M}$ is a linear map
\begin{equation}
\Im : T{\cal M}\oplus T^*{\cal M}\rightarrow
T{\cal M}\oplus T^*{\cal M},\:\mbox{\rm such that } 
\Im^2 = - \mbox{\bf 1}.
\end{equation}
This extends to a generalised complex structure when $\Im$ can be
defined consistently over ${\cal M}$ leading to integrability
conditions similar to the ordinary one\cite{HitchinGCY,Gualtieri}.
The dimension of the manifold ${\cal M}$ must be even. Trivial examples
are ordinary complex structure $I$, together with its transpose on
$T^*{\cal M}$:
\begin{equation}\label{trivialC}
\Im_I = \begin{pmatrix} I &0\\ 0 &-I^t\end{pmatrix},
\end{equation}
and symplectic structure, along with its inverse:
\begin{equation}\label{trivialS}
\Im_\omega = \begin{pmatrix} 0 &-\omega^{-1}\\ 
\omega &0\end{pmatrix}.
\end{equation}
In fact, a generalised complex structure is locally a product of
complex and symplectic structures\cite{Gualtieri}.

A generalised K\"ahler structure requires two commuting generalised 
complex structures $\Im_1$ and $\Im_2$, such that
\begin{equation}\label{gK}
\Im_1\,\Im_2 = -\, {\cal G} = -\,\begin{pmatrix} 0 &g^{-1}\\ 
g &0\end{pmatrix},
\end{equation}
where ${\cal G}$ is a positive definite generalised metric\footnote{
  In a more general situation, the generalised metric is a
  $B$-transform of the above\cite{Gualtieri}, in which case one
has the $B$-transforms of the $\Im$'s.} on $T{\cal M}\oplus
T^*{\cal M}$.  The motivation is clearly to extend the idea of a
hermitian K\"ahler manifold.

This structure first appeared in the physics literature in the work of
Gates {\em et al} \cite{GaHuRo}, where they analysed the requirement
for additional global supersymmetry for the sigma model describing a
string propagating in the presence of the Neveu-Schwarz $B$-field in a
target space $K$. This is a necessary requirement for spacetime
supersymmetry. They showed that the general form of the additional
supersymmetry variation of the (1,1) superfields $\Phi^i$ for the
coordinates, is of the form:
\begin{equation}\label{ssusyvar}
\delta \Phi^i = \epsilon^{+}f^i_{+j}D_{+}\Phi^{j} + 
\epsilon^{-} f^{i}_{-j}D_{-} \Phi^{j},
\end{equation}
where $D_{\pm}$ are the worldsheet super-covariant derivatives. The
tensors $f_{\pm}$ are a pair of almost complex structures on the
target manifold $K$ for the left- and right-movers respectively;
further, they have to satisfy the constraint
\begin{equation}\label{bihconst}
\nabla^{\pm}_{i}f^{j}_{\pm k} = \partial_{i}f^{j}_{\pm k} +
\Gamma^{j}_{\pm im}f^{m}_{\pm k} -\Gamma^{m}_{\pm ik}f^{j}_{\pm m}=0.
\end{equation}
In the above $\Gamma^{i}_{\pm jk}$ are the connections seen by the
left- and right movers
\begin{equation}\label{LRconn}
\Gamma^{i}_{\pm jk} = \Gamma^{i}_{jk} \pm g^{im}H_{mjk},
\end{equation}
where $\Gamma^{i}_{jk}$ are Levi-Civita connections derived from the
metric and $H=dB$.  This leads to what was called a bi-hermitian
geometry for $K$, namely distinct K\"ahler structures for the left-
and right-movers.  Gualtieri\cite{Gualtieri} has shown that the
bi-hermitian geometry of \cite{GaHuRo} can equivalently be written as
a generalised K\"ahler geometry.

\section{Generalised hyperK\"ahler geometry: strings in
a hyperK\"ahler manifold  with $H$-flux}

A generalised hyperK\"ahler structure naturally carries the idea of a
generalised K\"ahler structure one step further and requires six
generalised complex structures $\Im_a^\pm$, $(a=1,2,3)$. Recall that
an ordinary hyperK\"ahler structure has three complex structures $I_a$
($a=1,2,3$) transforming as an SU(2) triplet and three compatible
symplectic structures $\omega_a=gI_a$. The $\Im_a^\pm$'s satisfy the
following algebra that can be motivated from the hyperK\"ahler case:
\begin{eqnarray}\label{ghKalg}
\Im^+_a\,\Im^+_b &=& -\delta_{ab} + \epsilon_{abc}\Im^+_c,\nonumber\\
\Im^-_a\,\Im^-_b &=& -\delta_{ab} + \epsilon_{abc}\Im^+_c,\nonumber\\  
\Im^+_a\,\Im^-_b &=& -\delta_{ab}{\cal G} + \epsilon_{abc}\Im^-_c,\nonumber\\
\Im^-_a\,\Im^+_b &=& -\delta_{ab}{\cal G} + \epsilon_{abc}\Im^-_c,
\end{eqnarray}
where, $\epsilon_{abc}$ is a totally antisymmetric symbol and ${\cal
  G}$ is as in Eq.\refb{gK}.

As we have seen above, for a string moving in a manifold with the
metric $g$ in the presence of a $H$-flux, there is a torsion term that
modifies the connection for the left- and right-movers. In case of
(4,4) supersymmetry on the worldsheet, there are three complex
structures $f^{\pm}_{a}$ ($a=1,2,3$) in each sector.  The generalised
hyperK\"ahler structure is obtained straightforwardly by following
from the prescription of Ref.\cite{Gualtieri}:
\begin{eqnarray}\label{ghKs}
\Im^+_{a} &=& \begin{pmatrix} 
\frac{1}{2}\left(f^{+}_{a}+f^{-}_{a}\right)
& \frac{1}{2}\left(f^{+}_{a}-f^{-}_{a}\right)g^{-1}\\
\frac{1}{2}g\left(f^{+}_{a}-f^{-}_{a}\right) 
& -\frac{1}{2}\left(f^{t+}_{a}+f^{t-}_{a}\right)
\end{pmatrix},\nonumber\\
{}&{}&{}\nonumber\\
\Im^-_{a} &=& \begin{pmatrix} 
\frac{1}{2}\left(f^{+}_{a}-f^{-}_{a}\right)
& \frac{1}{2}\left(f^{+}_{a}+f^{-}_{a}\right)g^{-1}\\
\frac{1}{2}g\left(f^{+}_{a}+f^{-}_{a}\right) 
& -\frac{1}{2}\left(f^{t+}_{a}-f^{t-}_{a}\right)
\end{pmatrix},
\end{eqnarray}
where $f^{t\pm}_{a}$ denote the transpose of $f^{\pm}_{a}$. It is easy
to see that these satisfy the algebra \refb{ghKalg}.


\section{Generalised hyperK\"ahler geometry of the 
NS5-brane}

We will show that the Neveu-Schwarz 5-brane solution found in
Ref.\cite{CaHaSt} provides an example of generalised hyperK\"ahler
geometry. The NS5-brane was given as a soliton solution in the
supergravity approximation, however, at the same time a worldsheet
conformal field theory description established it as an exact solution
of string theory. Suppose the NS5-brane extends along the coordinates 
$(x^1\cdots x^5)$. The space labelled by $(x^6\cdots x^9)$ is the
transverse space $K$. In the supergravity approximation, the metric 
$g$ of $K$ is such that in the near horizon limit the geometry is
that of a cylinder with an $S^3$ base. There is an $H$-flux through the
$S^3$ and also a linear dilaton along the length of the cylinder.
Explicitly, the background is given by:
\begin{eqnarray}\label{nsfive}
g_{ij} &=& e^{2\phi} \delta_{ij},\quad i,j\ldots=6,\cdots,9,
\nonumber\\
H_{ijk} &=& -{\epsilon_{ijk}}^m\partial_m\phi,\\ 
\nabla^2e^{2\phi} &=& 0,
\end{eqnarray}
where $\phi$ is the dilaton field. It was also shown that there is
an exact (4,4) superconformal field theory on the worldsheet, for 
details see \cite{CaHaSt}. 

As mentioned earlier, the torsion term modifies the connection.
Therefore, the left- and right-movers perceive different hyperK\"ahler
structures as follows\cite{CaHaSt}:
\begin{equation}
\begin{array}{ll}
f^+_1 = \begin{pmatrix} i\sigma_{2} & 0 \\
                0 & -i\sigma_{2}\end{pmatrix},
&f^{-}_{1} = \begin{pmatrix} -i\sigma_{2} & 0 \\
                0 & -i\sigma_{2}\end{pmatrix},\\ 
{} &{}\\
f^{+}_{2} = \begin{pmatrix} 0 & 1 \\
                -1 & 0\end{pmatrix},
&f^{-}_{2} = \begin{pmatrix} 0 & -\sigma_{3} \\
                \sigma_{3} & 0\end{pmatrix},\\
{} &{}\\
f^{+}_{3} = \begin{pmatrix} 0 & i\sigma_{2} \\
                i\sigma_{2} & 0\end{pmatrix},
&f^{-}_{3} = \begin{pmatrix} 0 & -\sigma_{1}\\
                \sigma_{1} & 0\end{pmatrix}.
\end{array}
\end{equation}
The above, which are an extension of \cite{GaHuRo} to the
hyperK\"ahler case, may be called a bi-hyperK\"ahler structure. The
generalised hyperK\"ahler structure is then given by the
Eq.\refb{ghKs}. Clearly, these satisfy the algebra \refb{ghKalg}. Thus
we see that the (transverse space) of the NS5-brane of
Ref.\cite{CaHaSt} is a natural example of generalised hyperK\"ahler
manifold.



\section{Other proposals}

It is possible to find examples of generalised hyperK\"ahler
geometries, which are solutions to the string equations of motion at
order $\alpha'$.  To this end, consider the classical heterotic string
theory. Consider a spacetime of the form $\mbox{\bf R}^{6}\times K$
with no other background field. If $K$ is a Taub-Nut space or one of
the ALE spaces\cite{EgGiHa}, the lowest order equations of motion of
the Euclidean theory are satisfied, since the hyperK\"ahler space $K$
is a gravitational instanton. At order $\alpha'$ the Green-Schwarz
term modifies the equation of motion of the $B$ field:
\begin{equation}\label{greenschwarz}
dH = \alpha'\left(\mbox{\rm Tr}\,R\wedge R - \frac{1}{30}
\mbox{\rm Tr}\,F\wedge F\right).
\end{equation}
The usual practice is to turn on the gauge fields and identify the
gauge connection with the spin connection. This satisfies
Eq.\refb{greenschwarz}. 

Instead, one may consider a solution in which the gauge fields remain
trivial ($F=0$), but $H=\Omega_L(\sigma)$, where $\Omega_L(\sigma)$ is
the Lorentz Chern-Simons term for the spin connection $\sigma$ of the
hyperK\"ahler space $K$. The non-vanishing $H$-field will affect the
left- and right-movers differently, leading to different hyperK\"ahler
structures\footnote{This example was suggested by Ashoke Sen.} for
these. As we have seen, this is equivalent to a generalised
hyperK\"ahler structure.

Finally, we would like to propose a construction for a {\em compact}
generalised hyperK\"ahler manifold. It is in the context of the type
IIB string theory, in which there are two different 2-form fields
$B_{NS}$ and $B_{R}$ originating in the NS-NS and RR sectors
respectively. There is also an SL(2,{\bf Z}) duality symmetry under
which these transform as a doublet. Any of the SL(2,{\bf Z}) transform
of the $B$-fields can be combined with the metric to obtain a
generalised complex structure. Let us start with an elliptically
fibred K3 manifold. Now turn on an appropriate SL(2,{\bf Z}) transform
of the $B$-fields following the monodromy of the torus fibre, so that
these are consistent globally over the entire manifold.  Put in
another way, the singularities around which there are monodromies,
correspond to different $(p,q)$-5-branes, which are patched together
so that the transverse space is a {\em compact} generalised
hyperK\"ahler manifold.  This is in the spirit of the F-theory
construction of Ref.\cite{VafaF} in which $(p,q)$-7-branes are used.
This construction, if it can be made globally consistent, can also be
extended to other dimensions to obtain, for example, generalised
Calabi-Yau manifolds.

It should be interesting to check the details of the examples
mentioned in this section.

\vspace*{1ex}
\noindent{\bf Acknowledgement:} It is a pleasure to thank Arijit Dey,
Sunil Mukhi, Stefan Theisen and Ashoke Sen for discussions. BE would
like to thank the Harish-Chandra Research Institute and DG the Albert
Einstein Institute and the Tata Institute of Fundamental Research for
hospitality.


\end{document}